\documentclass[11pt]{article}

\usepackage[final]{acl}

\usepackage{times}
\usepackage{latexsym}
\usepackage[T1]{fontenc}
\usepackage[utf8]{inputenc}
\usepackage{microtype}
\usepackage[htt]{hyphenat}
\usepackage{inconsolata}
\usepackage{graphicx}
\usepackage{amsmath}
\usepackage{amssymb}
\usepackage{booktabs}
\usepackage{multirow}
\usepackage{xcolor}
\usepackage{subcaption}

\title{Incremental Pooled LLM Evaluation for Cost-Effective Retrieval Model Selection}

\author{
  Max Nelson,
  Hanoz Bhathena,
  Aviral Joshi,
  Saket Sharma \\
  JPMorganChase \\
  \texttt{max.nelson@jpmchase.com}
}

\begin{document}
\maketitle

\begin{abstract}
Selecting a retrieval model for a production RAG system requires reliable comparative evaluation, but obtaining relevance judgments at scale is expensive and difficult to repeat as new candidate systems arrive. We study pooled LLM evaluation, in which an LLM judges the union of documents retrieved by the current set of candidate systems, and the pool is then expanded incrementally as new systems are introduced by judging only the new documents they contribute. These judgments are reused to evaluate all systems on a common basis. We validate this approach on four retrieval benchmarks with 11 systems spanning dense, sparse, and hybrid configurations, and deploy it to compare 62 retrieval configurations for a financial news QA system. Pooled LLM rankings correlate strongly with gold-standard evaluation across datasets, and 97\% of pairwise system orderings are preserved once bootstrap uncertainty in the qrels is taken into account. In production, document overlap yields 65--80\% judgment reuse and up to 4.9x lower evaluation cost, allowing teams to benchmark new retrieval candidates without re-judging previously assessed documents. These results suggest pooled LLM evaluation is a practical and cost-effective workflow for incremental retrieval model selection in deployed systems.
\end{abstract}

\section{Introduction}
\label{sec:intro}

Retrieval-augmented generation (RAG) pipelines depend critically on the quality of the underlying retrieval model.
As new embedding models are released (OpenAI \texttt{text-embedding-3}, Cohere Embed v4, Nomic, Titan, etc.), practitioners face a recurring decision: \textit{which model gives the best retrieval quality for my domain?}

The gold standard for retrieval evaluation, pooled human relevance judgments, as in TREC~\cite{voorhees1998variations}, is expensive, slow, and difficult to repeat as new candidate systems arrive.
Evaluating $N$ systems independently with an LLM judge costs $O(N \cdot Q \cdot K)$ judgments (queries $\times$ top-$k$ documents per system), with substantial redundancy since different retrieval systems often surface overlapping document sets.

We evaluate \textit{pooled LLM evaluation}. We judge the \textit{union} of documents retrieved by all systems under test, then reuse these judgments to compute standard IR metrics for any system in the pool. As new systems are added to the set, pools grow based on only the new documents that those systems retrieve. 
This approach offers four advantages:

\begin{enumerate}
    \item \textbf{Cost amortization.} Shared documents are judged only once. Each additional system added to the pool incurs only marginal cost for its uniquely retrieved documents.
    \item \textbf{Ranking stability.} By definition, system-level P@$k$ and DCG@$k$ will always be invariant under pool expansion, and per-query pairwise orderings are preserved for recall@$k$, AP, and nDCG@$k$. Empirically, macro-averaged benchmark rankings remain highly stable as new systems are added.
    \item \textbf{Reduced pool bias.} Unlike classical TREC pooling, where unjudged documents are treated as non-relevant and systems with novel retrieval behavior are penalised~\cite{buckley2007bias}, pooled LLM judging can assess every newly retrieved document contributed to the pool. Teams can begin benchmarking with as few as two or three systems and add new ones over time, without systematic penalties for novel retrieval behavior due to unjudged documents.
    \item \textbf{Accuracy.} On four public IR benchmarks, pooled LLM rankings correlate strongly with gold-standard human judgments (nDCG@10 $\rho = 0.69$--$0.95$), and 74\% of pairwise ranking disagreements fall within the gold standard's own bootstrap sampling noise.

\end{enumerate}

\section{Related Work}
\label{sec:related}

\paragraph{LLM-as-Judge for IR Evaluation.}
Recent studies have demonstrated that LLM judges (GPT-4, Claude, etc.) produce relevance assessments that agree with human annotators at rates comparable to inter-annotator agreement~\cite{faggioli2023perspectives,thomas2024large}.

Recent TREC-related studies have explored LLM judges as part of the evaluation process and found that LLM-based relevance or support labels can yield system-level rankings that correlate well with rankings derived from human judgments~\cite{upadhyay2024v1,upadhyay2024v2}.

Our work builds on this foundation but addresses concerns of efficiency, bias, and ranking stability specifically as they pertain to industry practitioners. We focus on how to structure LLM evaluation to minimize cost while characterizing which aspects of ranking stability follow from metric definitions and which must be established empirically.

\paragraph{Test Collection Reusability.}
The reusability of pooled judgments is a classical topic in IR evaluation.
\citet{zobel1998reliable} and \citet{voorhees1998variations} established that TREC-style pools produce stable system rankings even when new, unjudged systems are evaluated.
\citet{buckley2007bias} analysed the conditions under which \textit{pool bias} affects system comparison: when a new system retrieves documents absent from the original pool, those documents are unjudged and effectively treated as non-relevant, systematically disadvantaging novel retrieval strategies.

Our LLM-based approach provides a structural safeguard against this concern: because LLM judging is inexpensive, we judge \textit{all} documents retrieved by any system, ensuring that no relevant document goes unassessed.

\paragraph{Cost-Effective Evaluation.}
Prior work has explored reducing annotation cost through active selection~\cite{carterette2006minimal}, stratified sampling~\cite{pavlu2007practical}, and assessment budgeting.
Our approach is complementary: rather than selecting \textit{which} documents to judge, we judge all pooled documents but reuse judgments across systems, exploiting the natural overlap in retrieval results.

\section{Method}
\label{sec:method}

\subsection{Pooled LLM Evaluation}

Given a set of $N$ retrieval systems $\{S_1, \ldots, S_N\}$ and a query set $\mathcal{Q}$, we:

\begin{enumerate}
    \item \textbf{Retrieve:} For each system $S_i$ and query $q \in \mathcal{Q}$, obtain the top-$k$ ranked documents $R_i(q)$.
    \item \textbf{Pool:} Form the document pool $\mathcal{P}(q) = \bigcup_{i=1}^{N} R_i(q)$.
    \item \textbf{Judge:} For each unique $(q, d)$ pair where $d \in \mathcal{P}(q)$, obtain a graded relevance judgment from an LLM judge on a 3-point scale (0: not relevant, 1: partially relevant, 2: highly relevant).
    \item \textbf{Evaluate:} Compute standard IR metrics for each system $S_i$ using the shared pool judgments restricted to documents in $R_i(q)$.
\end{enumerate}

\noindent Critically, this process is incremental.
A team begins with an initial set of systems, builds the pool, and later adds system $S_{N+1}$ by retrieving with it and judging only the new documents $R_{N+1}(q) \setminus \mathcal{P}(q)$.
All previously computed judgments are reused exactly, and the new system is immediately comparable to all existing ones on an equal footing.

\subsection{Ranking Stability Under Pool Expansion}
\label{sec:stability}

We note the following stability properties, which follow directly from metric definitions under the assumption that previously assigned labels are not revised and each system's ranked list remains fixed.
P@$k$ and DCG@$k$ depend only on each system's own top-$k$ documents and their (frozen) relevance labels, so they are exactly invariant under pool expansion.
Recall@$k$, AP, and nDCG@$k$ each have a system-specific numerator (fixed) divided by a query-specific denominator shared across all systems and because all systems' scores for a given query are scaled by the same factor, pairwise orderings are preserved within each query.
When scores are macro-averaged across queries, the per-query scaling factors may differ, so benchmark-level invariance is not guaranteed in general. We evaluate this empirically in Section~\ref{sec:perm-stability}.

\subsection{Cost Analysis}

Let $K$ denote the top-$k$ retrieval depth.
Independent evaluation of $N$ systems requires up to $N {\cdot} |\mathcal{Q}| {\cdot} K$ judgments.
Under pooling, we judge only the unique $(q,d)$ pairs in the union pool: $C_{\text{pool}} = \sum_q |\bigcup_i R_i(q)|$.
The \textit{reuse rate}, $1 - C_{\text{pool}} \,/\, (N {\cdot} |\mathcal{Q}| {\cdot} K)$, measures the fraction of judgments shared across systems.
The marginal cost of adding system $N{+}1$ is $\Delta C_{N+1} = \sum_q |R_{N+1}(q) \setminus \mathcal{P}(q)|$, only its uniquely retrieved documents.

\section{Experimental Setup}
\label{sec:setup}

\subsection{Datasets}

We evaluate on four benchmarks with varying characteristics (Details in Table \ref{tab:datasets}).
FiQA~\cite{maia2018www}, TREC-COVID~\cite{voorhees2021treccovid}, and Natural Questions (NQ)~\cite{kwiatkowski2019natural} are sourced from the BEIR zero-shot retrieval benchmark~\cite{thakur2021beir}; FinRAGBench-V~\cite{zhao2025finragbench} is a recent financial-domain multimodal RAG benchmark. We employ a layout-aware parser to extract text, tables, and figures from each document page. Extracted tables and figures are independently processed by a multimodal large language model (MLLM) to generate natural language descriptions, following the approach described in \citet{bhathena2026}. The resulting textual descriptions are then interleaved with the extracted text spans in natural reading order, preserving the original spatial layout of the document. To retain modality provenance, table and figure descriptions are enclosed in XML tags that explicitly mark their source modality.

\begin{table}[t]
\centering
\footnotesize
\setlength{\tabcolsep}{3.5pt}
\begin{tabular}{lcccc}
\toprule
& \textbf{FiQA} & \textbf{T-COVID} & \textbf{NQ} & \textbf{FinRAG-V} \\
\midrule
Queries & 648 & 50 & 500\textsuperscript{*} & 536 \\
Corpus size & 57K & 171K & 2.68M & 10.9K \\
Pool size & 249,865 & 18,870 & 188,903 & 192,783 \\
$n$ qrels & 1,706 & 66,336 & 3,452 & 606 \\
Rel/query (avg) & $\sim$2.6 & $\sim$1,327 & $\sim$1.0 & $\sim$1.1 \\
Relevance scale & Binary & 0-2 & Binary & Binary \\
Systems & 11 & 11 & 11 & 11 \\
\bottomrule
\end{tabular}
\caption{Dataset characteristics. \textsuperscript{*}First 500 of 3,452 test queries used for NQ.}
\label{tab:datasets}
\end{table}

\subsection{Retrieval Systems}

We evaluate systems spanning five embedding model families in three retrieval configurations:
\textbf{dense} (embedding-only retrieval with cosine similarity),
\textbf{sparse} (BM25 keyword retrieval), and
\textbf{hybrid} (reciprocal rank fusion of dense and sparse results).
Embedding models: OpenAI \texttt{text-embedding-3-large}, Cohere Embed v4, Amazon Titan Embed v2, Nomic \texttt{embed-text-v1.5}, and Nomic \texttt{embed-text-v2-moe}.
All embeddings are projected to 256 dimensions via Matryoshka truncation where supported.
This yields 5 dense systems, 5 hybrid systems, and 1 sparse (BM25) baseline per dataset.

\subsection{LLM Judge}
\label{sec:judge}

We use GPT-4.1 (\texttt{gpt-4.1-2025-04-14}) as the relevance judge, called at temperature 0.0 for near deterministic outputs.
Each $(q, d)$ pair is judged independently in a single-turn prompt (no in-context examples, no system-level calibration), ensuring that judgments are reproducible and order-independent.

\paragraph{Prompt Design.}
The prompt (Appendix~\ref{sec:prompt}) instructs the judge to assign a graded relevance score on a 3-point scale:
\textbf{0} (not relevant - nothing in the document addresses the query),
\textbf{1} (partially relevant - peripheral information is present but the core intent remains unanswered), and
\textbf{2} (fully relevant - the document provides a complete answer).
We adopt a 3-point scale rather than binary to capture the distinction between documents that are tangentially related versus directly responsive, which is critical for metrics like nDCG that weight graded relevance.
The prompt requires a brief chain-of-thought (``thought'' field, $\leq$2 lines) followed by the numeric score in a structured JSON output, encouraging the judge to reason before scoring and enabling post-hoc error analysis.
Documents are presented in a \texttt{Title: \ldots / Content: \ldots} format, giving the judge a consistent, structured view of each passage.

\paragraph{Judge Consistency.} To assess intra-judge reliability, we re-judged 1,000 randomly sampled $(q,d)$ pairs per dataset under identical conditions. Exact 3-point agreement was 93--96\% across the BEIR datasets ($\kappa = 0.84$--$0.90$), with $\geq$99.9\% agreement within $\pm$1 grade. Under binarization (scores 1--2 = relevant), agreement rises to 96--98\%. These results indicate that the judge is sufficiently stable to support reuse of pooled judgments.

\paragraph{Cross-Judge Agreement.}
To assess sensitivity to judge choice, we repeated the analysis with Claude Sonnet 4.6 using the same prompt. On 1,000 randomly sampled $(q,d)$ pairs per dataset, binarized agreement with GPT-4.1 was 84--92\% ($\kappa = 0.65$--$0.77$). To test whether these item-level differences affect system comparison, we re-judged the full TREC-COVID pool (18,870 documents) and recomputed rankings. Despite 27\% item-level disagreement, system rankings remained highly correlated between judges ($\rho = 0.89$ for nDCG@10, $0.92$ for recall@100, and $0.88$ for MAP), with 89\% pairwise agreement for both nDCG@10 and recall@100. Both judges identified the same top system on the two most important metrics. Precision@10 was less stable, consistent with its sensitivity to a small number of judgments per query. Overall, these results suggest that pooled evaluation is more robust at the system-ranking level than raw item-level agreement alone would imply.

\subsection{Evaluation Protocol}

For each dataset, all systems retrieve top-100 documents per query.
Documents are pooled and judged once.
We compute P@$k$, R@$k$, nDCG@$k$, and MAP against both \textbf{pseudolabels} (pooled LLM judgments, available for all datasets) and \textbf{gold qrels} (published human judgments, available for all four datasets; NQ and FinRAGBench-V qrels are very sparse with $\sim$1 relevant document per query).

\section{Results}
\label{sec:results}

\subsection{System-Level Rank Correlation}

Table~\ref{tab:correlation} reports Spearman and Kendall rank correlations between system rankings under pooled LLM judgments and gold-standard qrels.
Correlations are strongest on FiQA and NQ (nDCG@10 $\rho \geq 0.91$), where the LLM judge's broader relevance notion aligns well with the sparse qrels.
FinRAGBench-V shows similarly strong agreement ($\rho = 0.91$ for nDCG@10, $\rho = 0.94$ for R@100), consistent with its sparse qrels.
TREC-COVID yields lower but still significant correlations ($\rho = 0.64$--$0.82$), reflecting the gap between a general-purpose LLM and specialist assessors.

\begin{table*}[t]
\centering
\small
\begin{tabular}{lcccccccc}
\toprule
\textbf{Metric} & \multicolumn{2}{c}{\textbf{FiQA}} & \multicolumn{2}{c}{\textbf{TREC-COVID}} & \multicolumn{2}{c}{\textbf{NQ}} & \multicolumn{2}{c}{\textbf{FinRAG-V}} \\
& $\rho$ & $\tau$ & $\rho$ & $\tau$ & $\rho$ & $\tau$ & $\rho$ & $\tau$ \\
\midrule
nDCG@10 & .909 & .782 & .691 & .564 & .945 & .818 & .909 & .782 \\
P@10    & .918 & .818 & .636 & .527 & .864 & .745 & .936 & .818 \\
R@100   & .864 & .709 & .818 & .673 & .736 & .600 & .936 & .855 \\
MAP     & .855 & .673 & .736 & .527 & .718 & .600 & .882 & .745 \\
\bottomrule
\end{tabular}
\caption{Spearman ($\rho$) and Kendall ($\tau$) rank correlation between pooled LLM rankings and gold-standard qrels rankings. All correlations are statistically significant ($p < 0.05$).}
\label{tab:correlation}
\end{table*}

\subsection{Pairwise Rank Agreement}

The primary question for practitioners is whether pooled LLM evaluation preserves relative system orderings, i.e., if metrics calculated from qrel ground truths say system A is better than B, do pseudolabels agree?
We measure this over all $\binom{11}{2} = 55$ pairs per dataset:

\begin{table}[t]
\centering
\footnotesize
\setlength{\tabcolsep}{3pt}
\begin{tabular}{lcccc}
\toprule
\textbf{Metric} & \textbf{FiQA} & \textbf{T-COVID} & \textbf{NQ} & \textbf{FinRAG} \\
\midrule
nDCG@10 & 49 (89\%) & 43 (78\%) & 50 (91\%) & 49 (89\%) \\
P@10    & 50 (91\%) & 42 (76\%) & 48 (87\%) & 50 (91\%) \\
R@100   & 47 (85\%) & 46 (84\%) & 44 (80\%) & 51 (93\%) \\
MAP     & 46 (84\%) & 42 (76\%) & 44 (80\%) & 48 (87\%) \\
\bottomrule
\end{tabular}
\caption{Pairwise rank agreement (count and \%) between qrels and pseudolabel system rankings (55 pairs per dataset).}
\label{tab:pairwise}
\end{table}

FiQA and NQ show the strongest agreement (up to 91\%), while TREC-COVID, with its dense expert-annotated qrels and specialised biomedical domain, shows somewhat lower agreement (76--84\%).

\begin{figure}[t]
\centering
\includegraphics[width=\columnwidth]{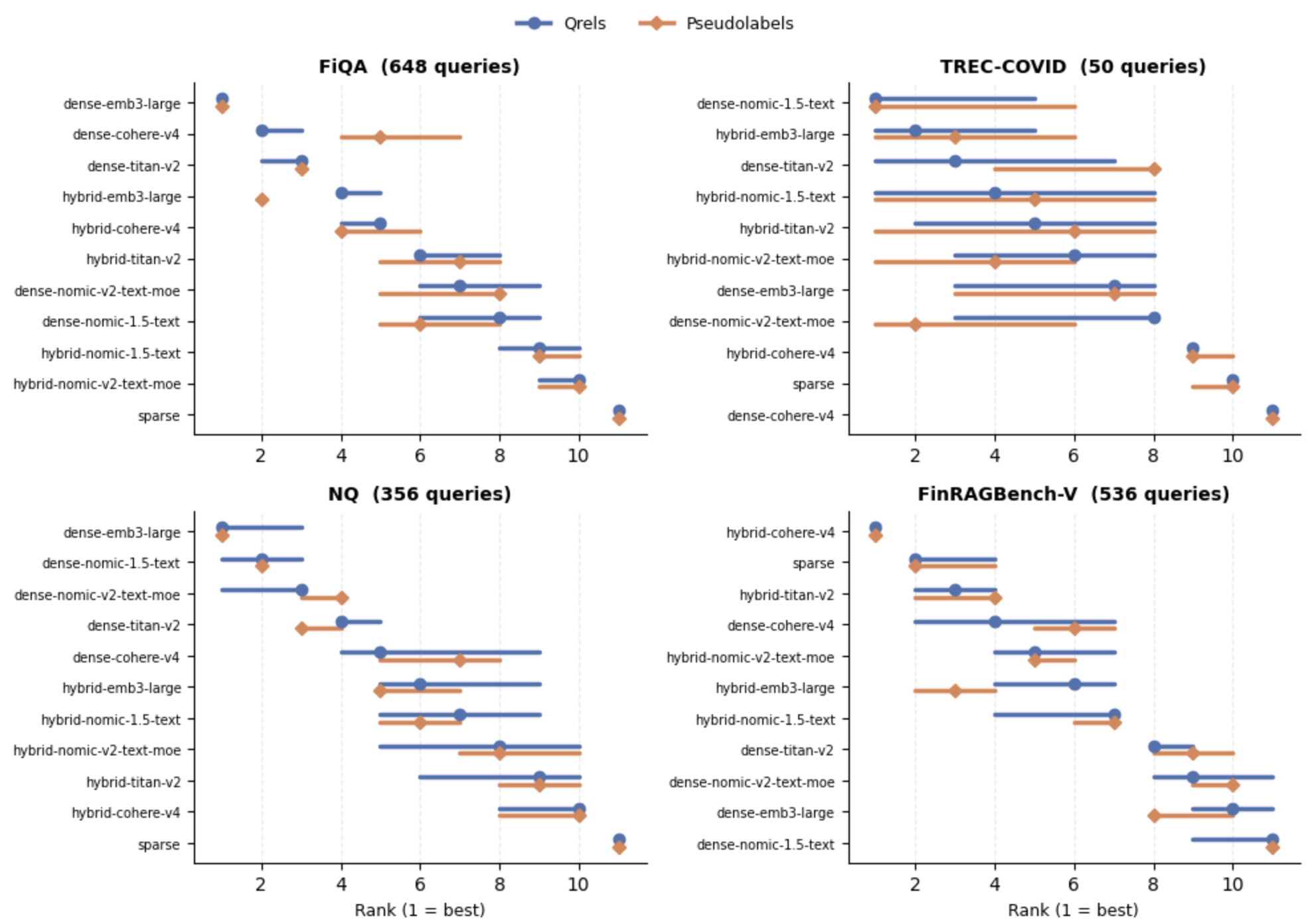}
\caption{Bootstrap 95\% rank confidence intervals for nDCG@10 across four datasets. Blue circles show qrels point ranks with CIs; orange diamonds show pseudolabel ranks with CIs. Most pseudolabel ranks fall within the qrels CI, indicating that ranking disagreements are within measurement noise.}
\label{fig:bootstrap}
\end{figure}

Critically, many of these disagreements fall within measurement noise.
To quantify measurement uncertainty in the gold standard itself, we compute \textit{bootstrap pairwise swap probabilities}: in each of $B = 10{,}000$ iterations we resample the $Q$ queries with replacement, recompute the macro-averaged metric for every system, and re-rank (full procedure in Appendix~\ref{sec:bootstrap-appendix}).
The swap probability for a pair $(A, B)$ is the fraction of iterations in which their ordering reverses.
A pair is \textit{qrels-uncertain} if swaps occur in $>$5\% of resamplings; a pairwise disagreement between qrels and pseudolabels is classified as \textit{within qrels noise} if the qrels themselves are uncertain about that pair.
For nDCG@10, 20 of 27 disagreements across all four datasets (74\%) fall within qrels noise; only 7 reflect genuine ranking differences.
Across all four metrics combined, 74 of 112 disagreements (66\%) are within noise on the three BEIR benchmarks; FinRAGBench-V shows the same pattern, with 4 of 6 nDCG@10 disagreements falling within qrels uncertainty.
Figure~\ref{fig:bootstrap} visualizes this: most pseudolabel ranks fall within the bootstrap 95\% rank confidence intervals of the qrels.
In practical terms, a developer using pooled LLM evaluation to decide between two retrieval models will reach the same conclusion as human evaluation in all but the most marginal cases.
Moreover, pooled LLM evaluation correctly identifies the rank-1 system on FiQA across all four metrics; the few disagreements on NQ and TREC-COVID occur only where the qrels margin between the top two systems is negligible (bootstrap swap probability $>$40\%).

\subsection{Cost Amortization}

Document overlap between retrieval systems produces substantial cost savings (Table~\ref{tab:cost}).
The marginal cost of adding a new system decreases as the pool grows, with each successive system requiring fewer novel judgments.

\begin{table}[t]
\centering
\footnotesize
\setlength{\tabcolsep}{3.5pt}
\begin{tabular}{lrrrc}
\toprule
\textbf{Dataset} & \textbf{Indep.} & \textbf{Pooled} & \textbf{Saved} & \textbf{Reuse} \\
\midrule
FiQA       & 712,800  & 249,865 & 462,935 & 65\% \\
TREC-COVID &  55,000  &  18,870 &  36,130 & 66\% \\
NQ         & 550,000  & 188,903 & 361,097 & 66\% \\
FinRAG-V   & 589,600  & 192,783 & 396,817 & 67\% \\
\midrule
\textbf{Total} & 1,907,400 & 650,421 & 1,256,979 & 66\% \\
\bottomrule
\end{tabular}
\caption{Judgment cost: independent ($N{\times}Q{\times}K{=}11{\times}Q{\times}100$) vs.\ pooled unique $(q,d)$ pairs. Pooling eliminates $\sim$1.26M redundant judgments across four datasets.}
\label{tab:cost}
\end{table}

\subsection{Ranking Stability Under Permuted Order}
\label{sec:perm-stability}

As discussed in Section~\ref{sec:stability}, pool expansion guarantees per-query order preservation but not benchmark-level invariance for macro-averaged metrics like nDCG@$k$.
To stress-test empirical stability, we simulate 500 random system-addition orderings per dataset: for each permutation we grow the pool from 1 to 11 systems and check whether any pairwise nDCG@10 ranking reverses at any step (Table~\ref{tab:perm-stability}).

\begin{table}[t]
\centering
\footnotesize
\setlength{\tabcolsep}{4pt}
\begin{tabular}{lclc}
\toprule
\textbf{Dataset} & \textbf{\% rev} & \textbf{Fragile pair} & $\boldsymbol{\Delta}$ \\
\midrule
FiQA       & 53.8 & nomic-1.5 / hyb-titan    & 1.4e-5 \\
           &      & hyb-nomic-1.5 / hyb-v2   & 1.2e-4 \\
NQ         & 0    & ---                      & --- \\
TREC-COVID & 0    & ---                      & --- \\
FinRAG-V   & 1.4  & emb3-large / titan-v2    & 5.9e-4 \\
\bottomrule
\end{tabular}
\caption{Permutation reversal search (nDCG@10, 500 random orderings). \% rev = orderings with $\geq$1 pairwise reversal during pool growth. $\Delta$ = final-pool nDCG@10 gap of the fragile pair.}
\label{tab:perm-stability}
\end{table}

In this test NQ and TREC-COVID exhibit \textit{zero} reversals across all 500 orderings.
On FiQA, 269 orderings (53.8\%) trigger a reversal, but this is confined to a single pair whose final nDCG@10 scores differ by only 0.000014 (dense-nomic-1.5 vs.\ hybrid-titan-v2); a second pair (hybrid-nomic-1.5 vs.\ hybrid-nomic-v2) reverses in only 2.4\% of orderings.
On FinRAGBench-V, 7 orderings (1.4\%) reverse the dense-emb3-large vs.\ dense-titan-v2 pair ($\Delta = 0.0006$).
In every case, the fragile pairs are the same ones that bootstrap rank CIs identify as statistically indistinguishable.
No pair with a meaningful score gap ($\Delta > 0.001$) reverses under any ordering on any dataset.

These results confirm that benchmark-level ranking instability under pool growth, while theoretically possible, is in practice confined to pairs whose scores are too close for any evaluation methodology to reliably distinguish.
\section{Analysis}
\label{sec:analysis}

\subsection{Calibration Bias}

Pseudolabel scores differ in absolute magnitude from qrels scores, the LLM's broader relevance notion inflates nDCG (mean $+$0.1\% to $+$51\% depending on dataset) while deflating recall ($-$17\% to $-$39\%), but this bias is \textit{approximately uniform across systems} on all four datasets (within-column std.\ $\leq$12.8 pp for nDCG, $\leq$11.1 pp for recall; Table~\ref{tab:calibration-bias} in Appendix~\ref{sec:calibration-appendix}).
Because the calibration offset does not vary meaningfully between retrieval configurations, it preserves ordinal system rankings.



\subsection{Sources of Disagreement}

Pairwise ranking disagreements are concentrated among systems with very small score differences. TREC-COVID shows the weakest agreement, consistent with its dense expert biomedical qrels and domain-specific relevance criteria. On the other three datasets, most disagreements occur among near-tied systems, and on FinRAGBench-V the remaining meaningful differences are concentrated in the \texttt{emb3-large} family. The same fragile pairs also appear in the pool-growth and bootstrap analyses, indicating that most disagreements reflect measurement-noise ties rather than substantive ranking errors.

\subsection{Pool Bias}
\label{sec:pool-bias}

Classical TREC pooling can penalize systems that retrieve documents absent from the pool, because unjudged documents are effectively treated as non-relevant~\cite{buckley2007bias}.
To quantify this risk, we ran a leave-one-family-out analysis: for each of six model families (five embedding families, plus sparse), we removed that family's unique contributions and treated them as non-relevant.

Pool bias is dataset-dependent.
FiQA and TREC-COVID were most affected (8 and 10 pairwise swaps, with maximum nDCG@10 penalties of 0.033 and 0.055, respectively), FinRAGBench-V showed limited impact (1 swap, max penalty 0.012), and NQ was effectively unaffected (0 swaps, max penalty 0.008), consistent with higher cross-family overlap.

By judging newly contributed documents whenever systems are added, pooled LLM evaluation avoids this classical failure mode.
In our experiments, this removed all 19 pairwise ranking errors introduced by classical pooling across the three affected datasets.

\section{Production Application}
\label{sec:deployment}

We applied pooled LLM evaluation to benchmark a set of embedding models for a production news question-answering system.
This section reports deployment experience, concrete cost figures, and practical lessons.

\subsection{Setting}

The production corpus consists of $\sim$300k news article pages indexed in OpenSearch, representing one month of news data that is licensed for use in internal generative AI applications. 
We constructed a benchmark of 303 queries from four sources: LLM-generated questions based on semantic article clusters (135), LLM-generated questions based on metadata-clustered articles by company, industry, and topic (91), templatic ``latest news'' questions for frequently mentioned entities (50), and real user queries from production API traffic, downsampled to remove near-duplicates (27).
Queries range from factual lookups (\textit{Who were the newly elected directors at TD Bank Group's 2025 annual meeting?}) to complex multi-hop multi-source questions (\textit{How is Stellantis addressing sales declines under its new CEO?}).
We further perturb these queries by converting them to keyword searches and to ``framed" queries, which specify output structure or formatting e.g. \textit{How is Stellantis addressing sales declines under its new CEO?} $\rightarrow$  
\textit{Summarize the recent changes at Stellantis under its new CEO and explain the company's strategies for addressing sales declines. Organize your response using section headings: Leadership Changes, Strategic Initiatives, and Sales Recovery Measures}.

\subsection{Systems Evaluated}

Using the pooled evaluation methodology, we benchmarked 62 retrieval configurations comprising
5 embedding models (OpenAI \texttt{emb3-large}, Cohere Embed v4, Nomic v1.5, Nomic v2-moe, MiniLM-L12-v2);
3 retrieval modes (dense, hybrid, sparse);
2--3 dimensionalities per model (256, 768, 1536);
3 query formulations per system (original, keyword-reduced, instruction-framed); and
variants of the top-performing system with and without query expansion.

\subsection{Cost Savings}

The final pool contains 766,350 unique $(q,d)$ judgments (all annotated by GPT-4.1).
Without pooling, evaluating the same set of experiments independently would have required 3,765,643 judgments, a 79.6\% reuse rate yielding a 4.9$\times$ cost reduction.
The pool was built incrementally over four weeks as new model candidates and query variants were added; at no point did previously computed judgments need to be re-run.

\subsection{Key Findings and Model Selection}

\paragraph{Production rankings.}
The top-5 systems by MAP@100 were all hybrid configurations:
(1) hybrid-emb3-large-768 (MAP 0.461),
(2) hybrid-emb3-large-256 (0.459),
(3) hybrid-nomic-1.5-768 (0.454),
(4) hybrid-nomic-v2-moe-768 (0.454),
(5) hybrid-cohere-v4-1536 (0.454).
Based on these results combined with latency and embedding cost considerations, we selected hybrid-emb3-large-256 for production deployment.

\paragraph{Additional findings.}
Higher dimensions improve dense retrieval (mean $\Delta$MAP = +0.029 for 256$\to$768/1536), but the gap narrows under hybrid retrieval (+0.012), suggesting BM25 fusion compensates for reduced embedding capacity.
Hybrid retrieval is also more robust to query formulation: keyword-reduced queries hurt dense models more ($\Delta$MAP = $-$0.013), and instruction-framed queries degrade all systems ($\Delta$MAP = $-$0.039) by diluting the retrieval signal.
Query expansion does not help on well-formed queries (MAP 0.424 vs.\ 0.459) but acts as a robustness mechanism against verbose formulations, nearly eliminating the framing-induced recall@100 penalty ($-$4.8pp $\to$ $-$1.1pp).

\subsection{Practical Lessons}

\paragraph{Incremental pool growth works.} We began with 5 systems and grew to 62 over four weeks. In our deployment, early conclusions (e.g., ``hybrid $>$ dense'') were not invalidated by later additions. This is qualitatively different from re-running an entire benchmark each time.

\paragraph{Cost is manageable.} At GPT-4.1 pricing (\$2/M input tokens, \$8/M output tokens), the 766K judgments cost approximately \$800 total.

\paragraph{Beyond model selection.} The same pool answered questions the team had not originally planned: ``Does dimensionality matter under hybrid?'' ``Are keyword queries harder?'' ``Does query expansion help on framed queries?''  Because all judgments are reusable, these analyses required zero additional LLM cost.

\paragraph{Judge stability.} Using a pinned model version (\texttt{gpt-4.1-2025-04-14}) at temperature 0 precludes drift by design within an evaluation campaign. If the judge model is later updated, selective re-judging of a sample can validate continuity.

\section{Discussion}
\label{sec:discussion}

\paragraph{Practical Implications.}
For production teams maintaining retrieval systems, pooled LLM evaluation enables a reusable incremental benchmarking workflow:
(1)~establish an initial pool with 3--5 candidate systems;
(2)~when a new model becomes available, retrieve with it and judge only the \textit{new} documents it contributes;
(3)~compare against all previously evaluated systems, with exact score reuse for previously judged documents and strong empirical ranking stability as the pool grows.

\paragraph{When This Method is Most Valuable.}
The approach is particularly suited to domains where:
(a) gold-standard relevance judgments do not exist or are prohibitively expensive to create;
(b) new embedding models are regularly evaluated (e.g., quarterly vendor updates);
(c) maintaining a stable historical benchmark is important for governance.

\section{Conclusion}
\label{sec:conclusion}

We presented pooled LLM evaluation as a practical framework for ongoing retrieval model selection.
In a production deployment, the method evaluated 62 retrieval configurations over four weeks at a total annotation cost of \$800, directly informing model selection for a financial news QA system.
Across four public benchmarks with 11 systems each, pooled LLM rankings correlate strongly with human judgments (nDCG@10 $\rho = 0.69$--$0.95$), and bootstrap analysis shows that 74\% of pairwise disagreements fall within the qrels' own sampling uncertainty, leaving only 7 meaningful nDCG@10 ranking differences across 220 system pairs.
Our empirical stress tests confirm that macro-averaged rankings remain stable as new systems are added to the pool.
Document overlap yields reuse rates of 65--80\%, making incremental evaluation economically viable as new embedding models arrive.

\section{Limitations}
\begin{itemize}
    \item \textbf{Tight clusters:} When systems are separated by $<$0.001 nDCG (e.g., the NQ top-2), neither qrels nor pseudolabels can reliably distinguish them, bootstrap swap probabilities exceed 40\% for both label sources. The method is designed for comparative model selection, not for resolving effectively tied systems.
    \item \textbf{Limited judge diversity:} System rankings are derived from a single primary judge (GPT-4.1). Full pool re-judging with Sonnet 4.6 on TREC-COVID confirms high system-level agreement ($\rho = 0.89$ nDCG@10; §\ref{sec:judge}), but this validation covers only two frontier-class LLMs on one dataset. Correlation and bias characteristics may differ for smaller, open-weight, or domain-specialized judges. Additionally, the frozen-model assumption means that if either provider deprecates the pinned checkpoint, re-judging a validation sample is necessary to confirm continuity.
    \item \textbf{System architecture scope:} All evaluated systems are single-stage retrievers (dense, sparse, or hybrid). We do not test reranking pipelines or cross-encoder architectures, where document overlap patterns and score distributions may differ substantially.
    \item \textbf{Benchmark-level stability:} Per-query pairwise invariance under pool expansion is exact for recall@$k$, AP, and nDCG@$k$ (Section~\ref{sec:stability}), but macro-averaged benchmark rankings can in principle shift when new systems change the relevance denominator. Our permutation stress tests (Section~\ref{sec:perm-stability}) show this is confined to pairs separated by $<$0.001, yet the guarantee remains empirical rather than universal.
\end{itemize}

\bibliography{custom}

\appendix
\section{Bootstrap Swap-Probability Details}
\label{sec:bootstrap-appendix}

This appendix gives the full procedure underlying the \textit{within qrels noise} classification reported in Section~\ref{sec:results}.
To quantify measurement uncertainty in the gold standard itself, we compute \textit{bootstrap pairwise swap probabilities}.
For each label source (qrels or pseudolabels), we construct a per-query score matrix $M \in \mathbb{R}^{Q \times N}$ where $M_{q,i}$ is the nDCG@10 score of system $i$ on query $q$.
In each of $B = 10{,}000$ bootstrap iterations, we resample $Q$ queries with replacement, compute the macro-averaged metric $\bar{s}_i = \frac{1}{Q}\sum_q M_{q,i}$ over the resampled set, and rank all $N$ systems (rank 1 = highest score; ties broken by minimum rank).
For each system pair $(A, B)$, the swap probability $p_{A \succ B}$ is the fraction of iterations in which $A$ is ranked above $B$.
A pair is classified as \textit{qrels-uncertain} if $\min(p_{A \succ B},\; 1 - p_{A \succ B}) > 0.05$, i.e., the less-favoured system wins in more than 5\% of query resamplings.
A pairwise disagreement between qrels and pseudolabels is then classified as \textit{within qrels noise} if the qrels themselves are uncertain about that pair under this criterion.
Under this criterion, pooled LLM evaluation correctly identifies the rank-1 system on FiQA across all four metrics; the few disagreements on NQ and TREC-COVID occur only where the qrels margin between the top two systems is negligible (bootstrap swap probability $>$40\%).

\section{Calibration Bias Details}
\label{sec:calibration-appendix}

Pooled LLM judgments exhibit consistent but dataset-dependent calibration bias relative to human qrels.
The LLM judge assigns graded relevance to a broader set of documents per query than human annotators.
This inflates nDCG numerators (more documents contribute gain) while simultaneously expanding the denominator of recall-based metrics (more documents are ``relevant,'' so any fixed-depth retrieval covers a smaller fraction).

\paragraph{FiQA.}
nDCG@10 pseudolabels score 29--74\% higher than qrels (mean +51\%), while recall@100 scores 33--45\% lower.

\paragraph{TREC-COVID.}
nDCG@10 bias is near zero (mean +0.1\%, range $-$17.5\% to +8.5\%).
However, recall@100 is 145--211\% higher under pseudolabels because TREC-COVID's exhaustive expert qrels identify $\sim$1,327 relevant documents per query, far more than the pool's 256 documents can cover.

\paragraph{NQ.}
nDCG@10 pseudolabels score 2--46\% higher (mean +17\%), while recall@100 scores 31--43\% lower.

\paragraph{FinRAGBench-V.}
nDCG@10 pseudolabels score 29--76\% higher (mean +43\%), while recall@100 scores 13--23\% lower (mean $-$17\%).
The pattern mirrors FiQA: sparse qrels ($\sim$1.1 rel/query) mean the LLM identifies many additional relevant documents.

\begin{table*}[t]
\centering
\small
\setlength{\tabcolsep}{3.5pt}
\begin{tabular}{l cc cc cc cc}
\toprule
 & \multicolumn{2}{c}{FiQA} & \multicolumn{2}{c}{TREC-COVID} & \multicolumn{2}{c}{NQ} & \multicolumn{2}{c}{FinRAG-V} \\
\cmidrule(lr){2-3} \cmidrule(lr){4-5} \cmidrule(lr){6-7} \cmidrule(lr){8-9}
System & nDCG & Recall & nDCG & Recall & nDCG & Recall & nDCG & Recall \\
\midrule
dense-cohere-v4 & $+29.1$ & $-43.1$ & $-17.5$ & $+183.1$ & $+11.0$ & $-42.8$ & $+29.9$ & $-22.6$ \\
dense-emb3-large & $+36.9$ & $-33.1$ & $+1.8$ & $+171.2$ & $+14.1$ & $-32.9$ & $+75.8$ & $-16.5$ \\
dense-nomic-1.5 & $+56.7$ & $-40.2$ & $+1.0$ & $+159.3$ & $+5.5$ & $-40.6$ & $+37.0$ & $-20.3$ \\
dense-nomic-v2 & $+55.1$ & $-39.5$ & $+8.5$ & $+171.0$ & $+1.9$ & $-42.4$ & $+52.3$ & $-20.4$ \\
dense-titan-v2 & $+38.8$ & $-44.5$ & $-4.9$ & $+152.4$ & $+12.8$ & $-42.1$ & $+42.9$ & $-18.8$ \\
hybrid-cohere-v4 & $+43.6$ & $-40.8$ & $-0.4$ & $+153.4$ & $+19.5$ & $-36.1$ & $+29.2$ & $-14.4$ \\
hybrid-emb3-large & $+55.6$ & $-35.3$ & $+0.9$ & $+153.3$ & $+21.6$ & $-30.7$ & $+53.9$ & $-12.6$ \\
hybrid-nomic-1.5 & $+58.9$ & $-38.1$ & $+1.3$ & $+151.8$ & $+21.9$ & $-36.5$ & $+33.0$ & $-16.6$ \\
hybrid-nomic-v2 & $+61.4$ & $-37.4$ & $+4.1$ & $+160.1$ & $+19.7$ & $-37.2$ & $+41.5$ & $-15.4$ \\
hybrid-titan-v2 & $+50.2$ & $-41.0$ & $+1.9$ & $+145.4$ & $+17.0$ & $-36.0$ & $+39.0$ & $-14.2$ \\
sparse (BM25) & $+74.1$ & $-36.4$ & $+4.1$ & $+172.7$ & $+46.4$ & $-32.2$ & $+39.4$ & $-12.8$ \\
\midrule
\textbf{Mean} & $+50.9$ & $-39.0$ & $+0.1$ & $+161.2$ & $+17.4$ & $-37.2$ & $+43.1$ & $-16.8$ \\
\textbf{Std.} & $\pm12.3$ & $\pm3.2$ & $\pm6.4$ & $\pm11.1$ & $\pm11.0$ & $\pm4.1$ & $\pm12.8$ & $\pm3.2$ \\
\bottomrule
\end{tabular}
\caption{Per-system calibration bias (\%$\Delta$: pseudolabel vs.\ qrels) for nDCG@10 and recall@100. The low standard deviation within each column confirms that bias is approximately uniform across systems, it differs in magnitude between datasets and metrics, but not between retrieval configurations. This uniformity preserves ordinal system rankings under pseudolabel evaluation.}
\label{tab:calibration-bias}
\end{table*}

\section{LLM Judge Prompt}
\label{sec:prompt}

The full prompt template used for graded relevance annotation is shown below.
Placeholders \texttt{\{passage\}} and \texttt{\{query\}} are filled at inference time.

\begin{small}
\begin{verbatim}
### Task Description: Determine the relevance
of the "DOCUMENT" to the "QUERY" provided below.

The goal is to assess the relevance of the
document with respect to a query based on the
document's "Title" and "Content".

### Relevance Scores:
0: Not Relevant - the document does not answer
   the query
1: Partially Relevant - the document contains
   peripheral information but the core intent
   remains unanswered
2: Fully Relevant - the document provides a
   full and complete answer to the query

### Output Guidelines:
1. Understand the Context
2. Apply the Relevance Criteria
3. Determine the Relevance Score (0-2)
4. Output as JSON:
   {"thought": "<2 lines max>",
    "relevance_score": <0, 1, or 2>}

DOCUMENT: {passage}
QUERY: {query}
\end{verbatim}
\end{small}

\end{document}